\documentclass[9pt,twocolumn,twoside]{pnas-new}
\templatetype{pnasresearcharticle}

\usepackage[numbers,sort&compress]{natbib}
\usepackage[capitalise]{cleveref}
\usepackage{graphicx}

\newcommand{\bm}[1]{\boldsymbol{#1}}

\DeclareMathOperator\Aut{Aut}
\newcommand\bW{{\mathbf{W}}}

\newcommand\cE{{\mathcal{E}}}
\newcommand\cV{{\mathcal{V}}}
\newcommand\tA{{\bm{\mathsf{A}}}}
\newcommand\tD{{\bm{\mathsf{D}}}}
\newcommand\tL{{\bm{\mathsf{L}}}}

\newcommand\tP{{\bm{\mathsf{P}}}}

\newcommand\grad{{\bm{\nabla}}}
\newcommand\transpose{{\!\mathsf{T}}}
\newcommand\prob{{\mathbb{P}}}

\newcommand\Methods{{(SI Numerical Methods)}}

\title{Stochastic cycle selection in active flow networks}

\author[a,1]{Francis G. Woodhouse}
\author[b]{Aden Forrow} 
\author[c]{Joanna B. Fawcett}
\author[b]{J\"orn Dunkel}

\affil[a]{Department of Applied Mathematics and Theoretical Physics, Centre for Mathematical Sciences, University of Cambridge, Wilberforce Road, Cambridge CB3 0WA, U.K.}
\affil[b]{Department of Mathematics, Massachusetts Institute of Technology, 77 Massachusetts Avenue, Cambridge MA 02139-4307, U.S.A.}
\affil[c]{Centre for the Mathematics of Symmetry and Computation, School of Mathematics and Statistics, The University of Western Australia, 35 Stirling Highway, Crawley, Perth, WA 6009, Australia}

\leadauthor{Woodhouse} 

\significancestatement{Nature often uses interlinked networks to transport matter or information. In many cases, the physical or virtual flows between network nodes are actively driven, consuming energy to achieve transport along different links. However, there are currently very few elementary principles known to predict the behavior of this broad class of non-equilibrium systems. Our work develops a generic foundational understanding of mass-conserving active flow networks. By merging previously disparate concepts from mathematical graph theory and physicochemical reaction rate theory, we relate the self-organizing behavior of actively driven flows to the fundamental topological symmetries of the underlying network, resulting in a new class of predictive models with applicability across many biological scales.}

\authorcontributions{All authors contributed at all stages of this work.}
\authordeclaration{The authors declare no conflict of interest.}
\correspondingauthor{\textsuperscript{1}To whom correspondence should be addressed. E-mail: F.G.Woodhouse@damtp.cam.ac.uk}


\begin{abstract}
Active biological flow networks pervade nature and span a wide range of scales, from arterial blood vessels and bronchial mucus transport in humans to bacterial flow through porous media or plasmodial shuttle streaming in slime molds. Despite their ubiquity, little is known about the self-organization principles that govern flow statistics in such non-equilibrium networks. Here we connect concepts from lattice field theory, graph theory, and transition rate theory to understand how topology controls dynamics in a generic model for actively driven flow on a network. Our combined theoretical and numerical analysis identifies symmetry-based rules that make it possible to classify and predict the selection statistics of complex flow cycles from the network topology. The conceptual framework developed here is applicable to a broad class of non-biological far-from-equilibrium networks, including actively controlled information flows, and establishes a new correspondence between active flow networks and generalized ice-type models.
\end{abstract}

\dates{This manuscript was compiled on \today}
\doi{}

\begin{document}

\maketitle
\thispagestyle{firststyle}
\ifthenelse{\boolean{shortarticle}}{\ifthenelse{\boolean{singlecolumn}}{\abscontentformatted}{\abscontent}}{}

\dropcap{B}iological flow networks, such as capillaries~\citep{Gazit1995_PRL}, leaf veins~\citep{Katifori2010_PRL} and slime molds~\citep{Alim2013_PNAS}, use an evolved topology or active remodeling to achieve near-optimal transport when diffusion is ineffectual or inappropriate~\citep{Banavar2000_PRL,Katifori2010_PRL,Corson2010_PRL,Nakagaki2000_Nature,Tero2010_Science}. Even in the absence of explicit matter flux, living systems often  involve flow of information currents along physical or virtual links between interacting nodes, as in neural networks~\citep{Bullmore2009_NatRevNeuro}, biochemical interactions~\citep{Jeong2000_Nature}, epidemics~\citep{PastorSatorras2015_RMP} and traffic flow~\citep{Garavello}. The ability to vary the flow topology gives network-based dynamics a rich phenomenology distinct from that of equivalent continuum models~\citep{Nakao2010_NatPhys}. Identical local rules can invoke dramatically different global dynamical behaviors when node connectivities change from nearest-neighbor interactions to the broad distributions seen in many networks~\citep{Watts1998_Nature,Barabasi1999_Science,Albert2002_RMP,Bianconi2009_PNAS}.
Certain classes of interacting networks are now sufficiently well understood  to be able to exploit their topology for the control of input--output relations~\citep{Nepusz2012_NatPhys,Menichetti2014_PRL}, as exemplified by microfluidic logic gates~\citep{Prakash2007_Science,Pearce2015_RSIf}.
However, when matter or information flow through a noisy network is not merely passive but actively driven by non-equilibrium constituents~\citep{Alim2013_PNAS}, as in maze-solving slime molds~\citep{Nakagaki2000_Nature}, there are no overarching dynamical self-organization principles known.  
In such an \emph{active network}, noise and flow may conspire to produce behavior radically different from that of a classical forced network. This raises the general question of how path selection and flow statistics in an active flow network depend on its interaction topology.

Flow networks can be viewed as approximations of a complex physical environment, using nodes and links to model intricate geometric constraints~\citep{Dias1986_JFM,DeArcangelis1986_PRL,Wu2012_LabChip}. These constraints can profoundly  affect matter transport~\citep{Misiunas2015_PRL,Illien2014_PRL,Benichou2015_JPhysA,Fuerstman2003_Langmuir}, particularly for active systems~\citep{Vicsek2012_PhysRep,Marchetti2013_RevModPhys,Dunkel2013_PRL} where geometric confinement can enforce highly ordered collective dynamics~\citep{Woodhouse2012_PRL,Wioland2013_PRL,Lushi2014_PNAS,Pearce2015_RSIf,Ravnik2013_PRL,Furthauer2012_NJP,Buhl2006_Science,Yates2009_PNAS,Paoluzzi2015_PRL,Bricard2013_Nature,Tjhung2012_PNAS}. 
In symmetric geometries like discs and channels, active flows can often be effectively captured by a single variable $\phi(t)$, such as angular velocity~\citep{Wioland2013_PRL,Wioland2016_NatPhys} or net flux~\citep{Yates2009_PNAS}, that tends to adopt one of two preferred states $\pm \phi_0$. External or intrinsic fluctuations can cause $\phi(t)$ to diffuse in the vicinity of, say, $-\phi_0$ and may occasionally trigger a fast transition to~$\phi_0$ and \textit{vice versa}~\citep{Wioland2016_NatPhys,Yates2009_PNAS}.
Geometrically coupling together many such confined units then results in a lattice field theory, reducing a non-equilibrium active medium to a discrete set of variables obeying pseudo-equilibrium physics, as was recently demonstrated for a lattice of bacterial vortices~\citep{Wioland2016_NatPhys}.
Below, we generalize this idea by constructing a generic lattice field model for an incompressible active medium flowing in an arbitrary network of narrow channels. Combining concepts from transition rate theory and graph theory, we show how the competition between incompressibility, noise and spontaneous flow can trigger stochastic switching between states comprising cycles of flowing edges separated by acyclic sets of non-flowing edges. As a main result, we find that the state transition rates for individual edges can be related to one another via the cycle structure of the underlying network, yielding a topological heuristic for predicting these rates in arbitrary networks.  We conclude by establishing a mapping between incompressible active flow networks and generalized ice-type or loop models~\citep{Baxter,Kondev1997_PRL,Batchelor1994_PRL}.

\section*{Model}

\subsection*{\boldmath Lattice $\phi^6$ field theory for active flow networks}

Our network is a set of vertices $v \in \cV$ connected by edges $e \in \cE$, forming an undirected loop-free graph $\Gamma$. 
(We use graph theoretic terminology throughout, where a loop is a single self-adjacent edge and a cycle is a closed vertex-disjoint walk.) To describe signed flux, we construct the directed graph $\hat\Gamma$ by assigning an arbitrary orientation to each edge. Now, let $\phi_e$ be the flux along edge $e$, where $\phi_e > 0$ denotes flow in the direction of the orientation of $e$ in $\hat\Gamma$ and $\phi_e < 0$ denotes the opposite. To model typical active matter behavior~\citep{Woodhouse2012_PRL,Wioland2013_PRL,Lushi2014_PNAS,Wioland2016_NatPhys}, we assume that fluxes either spontaneously polarize into flow states $\phi_e \approx \pm 1$, or adopt some other non-flowing mode $\phi_e \approx 0$.  We formalize this by imposing a bistable potential $V(\phi_e) = -\tfrac{1}{4} \phi_e^4 + \tfrac{1}{6}\phi_e^6$ on each flux variable. 
This form, of higher order than in a typical Landau theory, ensures that incompressible potential minima are polarized flows with every $\phi_e$ in the set $\{-1,0,+1\}$, rather than the continuum of fractional flow states that a typical $\phi^4$ potential would yield (SI Local Potential).
 
Incompressibility, appropriate to dense bacterial suspensions or active liquid crystals, is imposed as follows.
The net flux into vertex $v$ is $\sum_{e \in \cE} D_{ve} \phi_e$, where the discrete negative divergence operator $\tD = (D_{ve})$ is the $|\cV|\times|\cE|$ incidence matrix of $\hat\Gamma$ such that $D_{ve}$ is $-1$ if $e$ is directed out of $v$, $+1$ if $e$ is directed into $v$, and $0$ if $e$ is not incident to $v$~\citep{GodsilRoyle}.
Exact incompressibility corresponds to the constraint $\tD \Phi = 0$ on the global flow configuration $\Phi = (\phi_e) \in \mathbb{R}^{|\cE|}$. To allow for small fluctuations, modeling variability in the microscopic flow structure, we apply this as a soft constraint via an interaction potential $\propto |\tD\Phi|^2$.  The total energy $H(\Phi)$ of the active flow network then reads
\begin{align}
\label{eq:energy}
H(\Phi) =  \lambda \sum_{e \in \cE} V(\phi_e) + \tfrac{1}{2} \mu |\tD\Phi|^2,
\end{align}
with coupling constants $\lambda$ and $\mu$. 
This energy is comparable to that of a lattice spin field theory, but with interactions given by higher-dimensional quadratic forms akin to a spin theory on the vertices of a hypergraph (SI Model Detail).

\subsection*{Network dynamics}
Appealing to recent results showing that bacterial vortex lattices obey equilibrium-like physics~\citep{Wioland2016_NatPhys}, we impose that $\Phi$ obeys the overdamped Langevin equation
\begin{align}
\label{eq:langevin}
d\Phi = -\frac{\delta H }{\delta\Phi} dt + \sqrt{2\beta^{-1}} d\bW_t,
\end{align}
with $\bW_t$ an $|\cE|$-dimensional vector of uncorrelated Wiener processes and $\beta$ the inverse temperature. This stochastic dynamical system has a Boltzmann stationary distribution $\propto e^{-\beta H}$. The components of the energy gradient $\delta H/\delta \Phi$ in \cref{eq:langevin} are 
\begin{align}
\label{eq:func_deriv}
\left(\frac{\delta H}{\delta \Phi}\right)_{\!e}
&= - \lambda \phi_e^3 (1-\phi_e^2) + \mu (\tD^\transpose \tD \Phi)_{e}.
\end{align}
$\tD^\transpose \tD$ is the discrete Laplacian operator on edges, which is of opposite sign to the continuous Laplacian $\nabla^2$ by convention. The last term in~\cref{eq:func_deriv} arises in an otherwise equivalent fashion to how a bending energy $|\grad \psi|^2$  yields a diffusive term $\nabla^2 \psi$ in a continuous field theory.
On its own, this term damps non-cyclic components of the flow while leaving cyclic components untouched; these components' amplitudes would then undergo independent Brownian walks were they not constrained by the $\phi^6$ component of $V$ (SI Model Detail).

We now characterize the behavior of this model on a variety of forms of underlying graph $\Gamma$. For clarity, in addition to our prior assumption that $\Gamma$ is loop-free (which simplifies definitions and is unimportant dynamically since loops decouple; SI Model Detail) we will focus on connected, simple graphs $\Gamma$, though multiple edges are not excluded \textit{per se} (Fig.~S1). In what follows, we work in the near-incompressible regime $\mu \gg \lambda$ before discussing the strictly incompressible limit $\mu \to \infty$ below.

\section*{Results}

\begin{figure*}[t]
\includegraphics[width=\textwidth]{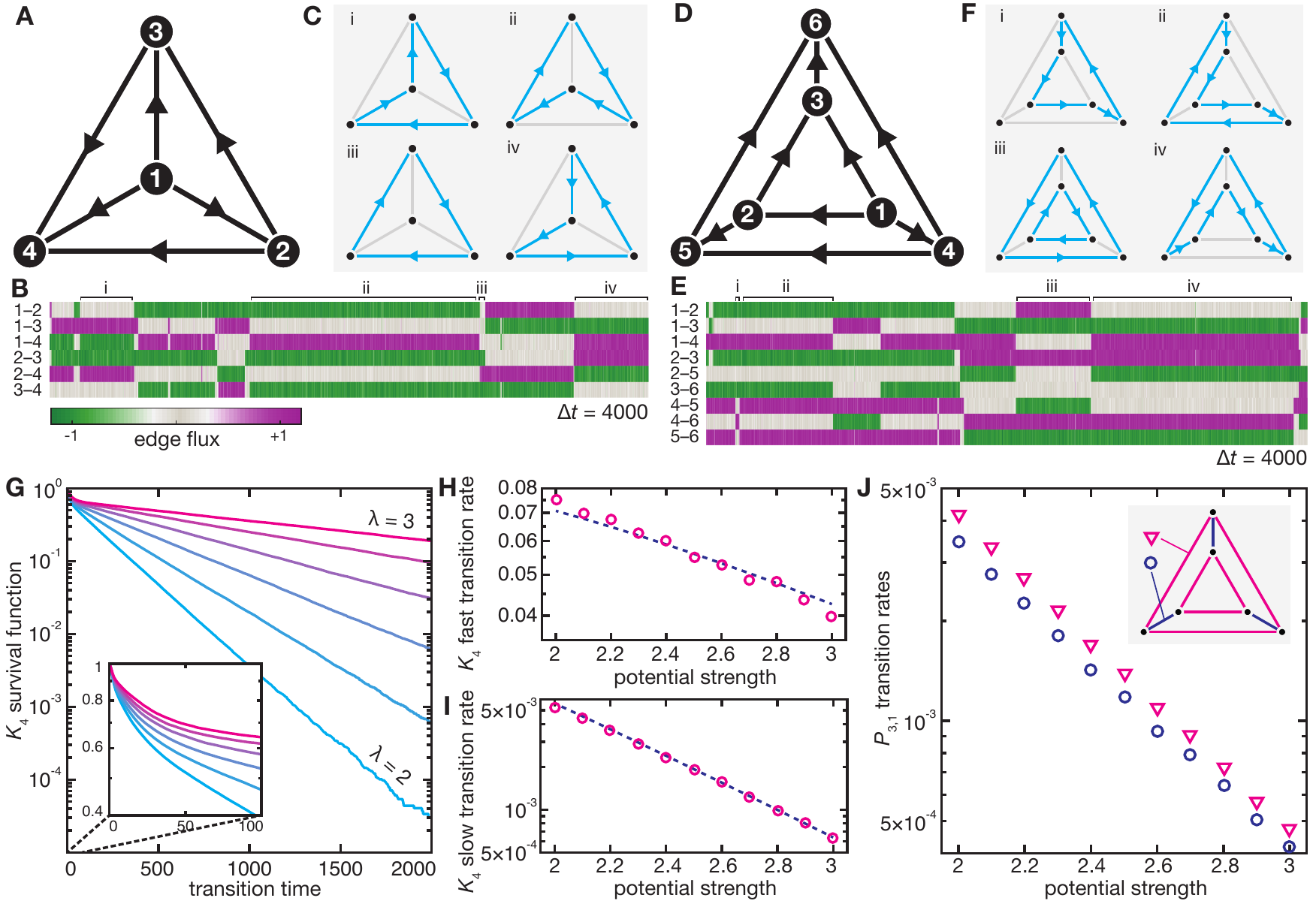}
\caption{Noise and activity cause stochastic cycle selection.
(A--C)~Flux--time traces (B) for each edge of the complete graph on four vertices, $K_4$. Edge orientations are as in (A). The sub-diagrams in (C)(i--iv) exemplify the flow state in the corresponding regions of the trace.
Parameters $\lambda = 2.5$, $\mu = 25$, $\beta^{-1} = 0.05$.
(D--F)~As in (A--C), but for the generalized Petersen graph $P_{3,1}$. The same switching behavior results, but now with more cycle states.
See also SI Movie 1.
(G)~Survival function $S(t) = \prob(T > t)$ of the transition waiting time $T$ for an edge in $K_4$, at regularly-spaced values of $\lambda$ in $2 \leqslant \lambda \leqslant 3$ with $\mu = 25$, $\beta^{-1} = 0.05$. Log-scaled vertical; straight lines imply an exponential distribution at large $t$. Inset: $S(t)$ at small $t$ with log-scaled vertical, showing non-exponential behavior.
(H,I)~Slow and fast edge transition rates in $K_4$, with parameters as in (G). Circles are from fitting $T$ to a mixture of two exponential distributions, lines show best-fit theoretical rates $k \propto \lambda \exp(-\beta \Delta H)$ with $\Delta H$ calculated for transitions between 3- and 4-cycles (SI Energy Barriers).
(J)~Transition rate $k = \langle T \rangle^{-1}$ for each set of equivalent edges in $P_{3,1}$, as per the key, as a function of $\lambda$, with $\mu = 25$, $\beta^{-1} = 0.05$. Log-scaled vertical shows exponential dependence on $\lambda$.
\label{fig:switching}}
\end{figure*}

\begin{figure*}[t]
\includegraphics[width=\textwidth]{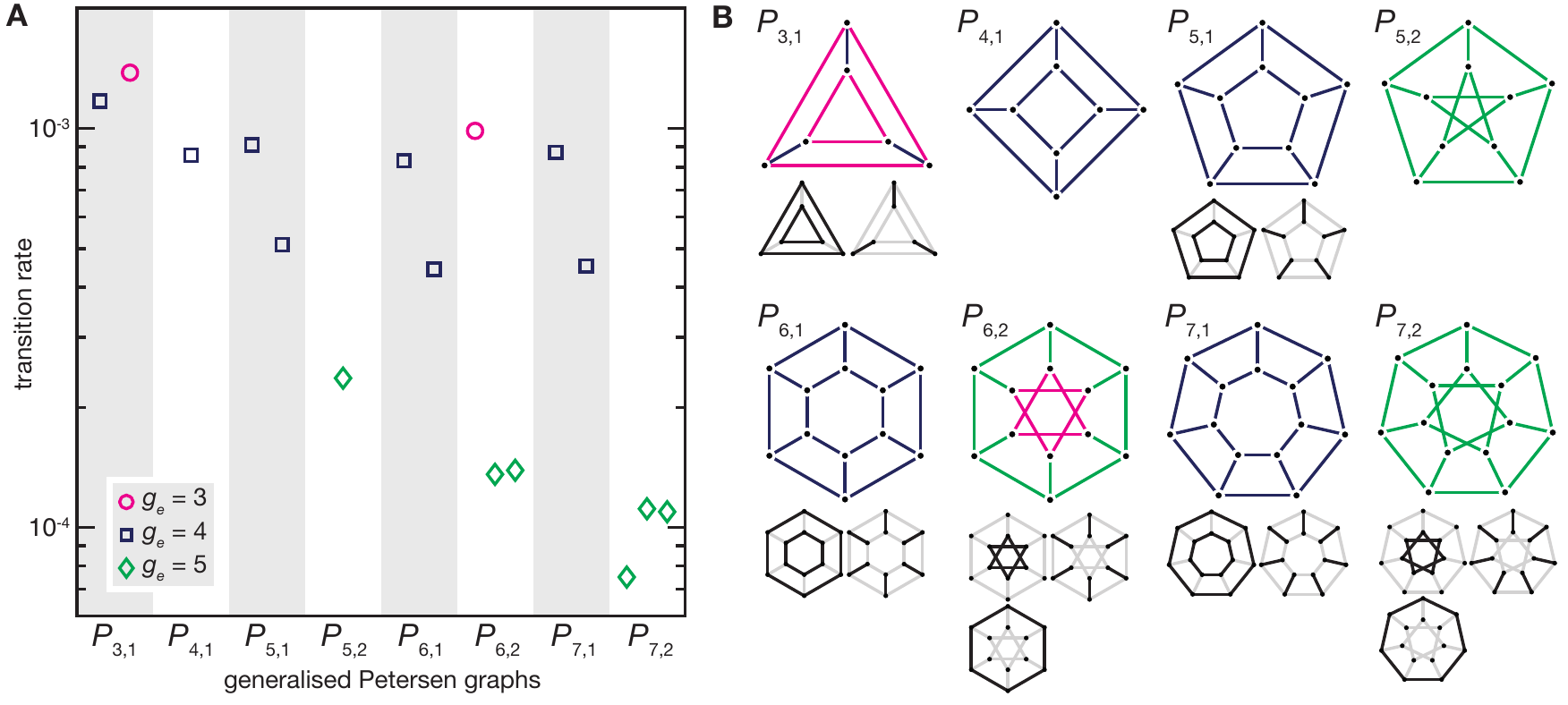}
\caption{Transition rates in highly symmetric graphs are determined by cycle structure.
(A)~Transition rate for edges in the first eight generalized Petersen graphs with $\lambda = 2.5$, $\mu = 25$, $\beta^{-1} = 0.05$. The rate was determined for each edge, then averaged within classes of equivalent edges. Symbols denote the rate for each class, categorized by $e$-girth $g_e$ as in the key. The range of computed rates within each class is smaller than the symbols.
(B)~The graphs in (A) with their edge equivalence classes when more than one exists. Edges colors denote $g_e$ as in (A). Observe that identical $e$-girth does not  imply equivalence of edges.
\label{fig:petersen}}
\end{figure*}
 
\subsection*{Stochastic cycle selection}
The combination of energy minimization and noise leads to \emph{stochastic cycle selection}~(SI Movies).
A local energy minimum comprises a maximal edge-disjoint union of unit-flux cycles: edge fluxes seek to be at $\pm 1$ if possible subject to there being zero net flux at every vertex, leading to states where the non-flowing edges contain no cycles (that is, they form a \emph{forest}, or a union of trees).
However, noise renders these states only metastable and induces random switches between them.
\Cref{fig:switching}A--F depicts flow on the 4-vertex complete graph~$K_4$ (\cref{fig:switching}A--C) and the generalized Petersen graph~$P_{3,1}$ (\cref{fig:switching}D--F)---the tetrahedron and triangular prism, respectively---where we have integrated \cref{eq:langevin} to yield flux--time traces of each edge \Methods.
The coordinated switching of edges between states of mean flux at $-1$, $0$ and $+1$ leads to random transitions between cyclic states, as illustrated. Note that the more flowing edges a state has, the lower its energy and therefore the longer-lived that state will be; thus in $K_4$, for example, 4-cycles, which are global minima, persist longer than 3-cycles~(\cref{fig:switching}B,C).

A graph possessing an Eulerian cycle---a non-repeating tour of all edges starting and ending at one vertex---has global energy minima with all edges flowing, which exists if and only if all vertices are of even degree. By contrast, a graph possessing many vertices of odd degree will have minimum energy states with non-flowing edges, because edges flowing into and out of such a vertex pair up to leave an odd number of $0$-flow edges. Such `odd' networks are particularly interesting dynamically as they are more susceptible to noise-induced state switches than graphs with even degree vertices (SI Model Detail).
For this reason, we specialize from now on to \emph{cubic} or \emph{3-regular graphs} where all vertices have degree three. 

\subsection*{Waiting times and graph symmetries}
The cycle-swapping behavior can be quantified by the distribution of the waiting time for an edge to transition between states in $\{-1,0,1\}$. 
For some edges, dependent on $\Gamma$, this distribution will be identical: the interactions in the energy (\ref{eq:energy}) are purely topological, with no reference to an embedding of $\Gamma$, implying that only topological properties---in particular, graph symmetries---can influence the dynamics.
Symmetries of a graph $\Gamma$ are encoded in its automorphism group $\Aut(\Gamma)$, whose elements permute vertices and edges while preserving incidence and non-incidence~\citep{GodsilRoyle}.
Two edges will then follow identical state distributions if one can be mapped to the other by some element of $\Aut(\Gamma)$ (SI Automorphic Equivalence); this determines an equivalence relation on $\cE$.
In $K_4$, every vertex is connected to every other, so $\Aut(K_4) = S_4$. This means any edge can be permuted to any other---the graph is \emph{edge transitive}---so all edges are equivalent and may be aggregated together.
To quantify cycle swapping in $K_4$, we numerically determined the distribution of the waiting time for an edge to change its state between $-1$, $0$ and $+1$ \Methods.
The resultant survival function $S(t) = \prob(T > t)$ for the transition waiting time $T$ of any edge in $K_4$ lengthens with increasing flow polarization strength $\lambda$ (\cref{fig:switching}G), and is well approximated by a two-part mixture of exponential distributions.

\subsection*{Transition rate estimation}
Reaction-rate theory explains the form of the waiting time distribution~\cite{Hanggi1990_RevModPhys}.
In a system obeying damped noisy Hamiltonian dynamics such as ours, a transition from one local energy minimum to another, respectively $\Phi_a$ and $\Phi_c$, will occur along a one-dimensional submanifold crossing a saddle point $\Phi_b$. The waiting time $T_{ac}$ for this transition to occur is then distributed approximately exponentially, with rate constant $k_{ac} = \langle T_{ac} \rangle^{-1}$. For a Hamiltonian that is locally quadratic everywhere, this results in the generalized Arrhenius law~\citep{Hanggi1990_RevModPhys}
\begin{align}
\label{eq:transition_rate_theory}
k_{ac} \propto \left[- \nu^{(b)}_1 \frac{\prod_{i=1}^N \nu^{(a)}_i}{\prod_{i=2}^N \nu^{(b)}_i} \right]^{1/2} \exp(-\beta \Delta H_{ab}),
\end{align}
where $\Delta H_{ab} = H(\Phi_b) - H(\Phi_a)$ is the transition energy barrier, 
and $\nu^{(a)}_i, \nu^{(b)}_i$ are the eigenvalues of the Hessian $\delta^2 H/\delta\Phi^2$ with $\nu^{(b)}_1 < 0$  the unstable eigenvalue at the saddle point.
The reverse transition time $T_{ca}$ obeys another exponential distribution with equivalent rate $k_{ca}$ dependent on the energy barrier $\Delta H_{cb}$. 
Therefore, the aggregated distribution of the waiting time $T$ for the system to change its state between either minimum is a mixture of two exponential distributions weighted by the equilibrium probabilities of the system to be found in each state.
For $K_4$, almost all transitions should be between 3-cycles and 4-cycles: each Eulerian subgraph is a 3-cycle or a 4-cycle, with 4-cycles being global minima, and direct transitions between different 4-cycles have a large enough energy barrier to be comparatively rare. Thus, we expect $K_4$ to exhibit a two-part mixture with a slow rate $k_{43}$ from a 4- to a 3-cycle and a fast rate $k_{34}$ from a 3- to a 4-cycle.
\Cref{fig:switching}H,I shows $k_{43}$ and $k_{34}$ for $K_4$ at a range of values of $\lambda$, as determined by maximum likelihood estimation on simulation data.
Our non-quadratic potential means these rates are not precisely determined by \cref{eq:transition_rate_theory}, but it does suggest an Arrhenius-type dependence $k_{ac} \propto \lambda \exp(-\beta \Delta H_{ab})$.
Computing the energy barriers (SI Energy Barriers)~and fitting the proportionality constant for each of $k_{34}$ and $k_{43}$ then gives excellent fits to the data, confirming our hypothesis (\cref{fig:switching}H,I).

The complete graph $K_4$ has as much symmetry as is possible on four vertices. This is unusual; most graphs have multiple classes of equivalent edges.
Though $P_{3,1}$ (\cref{fig:switching}D) is vertex transitive, in that any vertex can be permuted to any other by its automorphism group $\Aut(P_{3,1}) = D_6 \times C_2$, it is \emph{not} edge transitive. Instead, the edges split into two equivalence classes (\cref{fig:switching}J, inset), one containing the two triangles and the other containing the three edges between them.
The waiting times then cluster into two distinct distributions according to these two classes.
However, when more than two inequivalent minima exist, as they do for $P_{3,1}$, the potential transitions rapidly increase according to the combinatorics of the mutual accessibility between these minima. 
On $P_{3,1}$ there is potentially one rate for each pairwise transition between 4-, 5- and 6-cycles, leading to a mixture of six or more exponentials for the waiting time distribution which cannot be reliably statistically distinguished without large separations of time scales.
Instead, we compute the transition rate $k = \langle T \rangle^{-1}$ for each set of equivalent edges. The rates decay exponentially with $\lambda$ (\cref{fig:switching}J), consistent with transitions obeying \cref{eq:transition_rate_theory}. But why does one set of edges transition slower on average than the other? We shall now explore this question for both highly symmetric and totally asymmetric graphs.

\begin{figure}[t]
\centering
\includegraphics[width=\columnwidth]{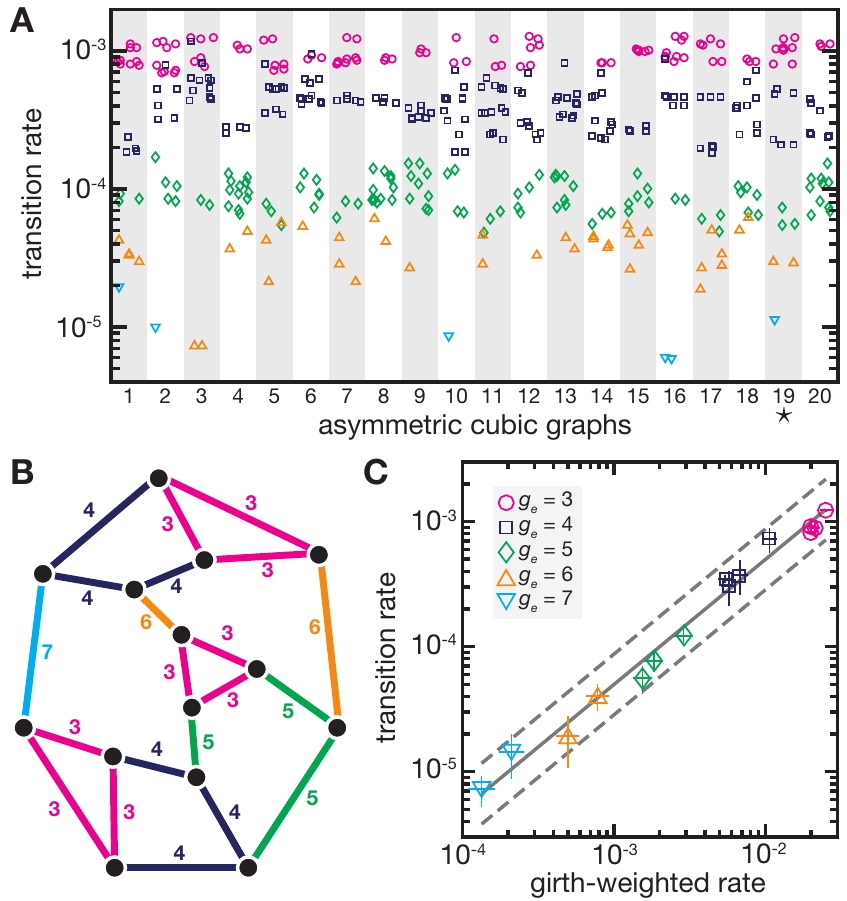}
\caption{Cycle structure determines edge transition rates in asymmetric graphs.
(A)~Transition rate for each edge in 20 random asymmetric bridgeless cubic graphs on 21 edges \Methods. Markers denote $e$-girth $g_e$ as per the key in (C). $\lambda = 2.5$, $\mu = 25$, $\beta^{-1} = 0.05$.
(B)~One of the graphs in (A), corresponding to the marked column ($\star$). Edges colored and labelled according to $g_e$. All 20 graphs are shown in Fig.~S4. See also SI Movie 2.
(C)~Transition rates $k$ from (A) binned by girth-weighted rate~$G$,  using best-fit value $\alpha = 1.31$, with markers denoting $g_e$ as per the key. Horizontal error bars are range of marker position over 95\% confidence interval in~$\alpha$, vertical error bars are $\pm 1$ standard deviation in $k$ within each group. Solid line is best fit $k = \gamma G$, dashed lines are 95\% prediction intervals on $k$ with $\alpha$ fixed.
\label{fig:asymmetric}}
\end{figure}

\subsection*{Edge girth determines rate band structure}
Global symmetries and local graph structure play distinct roles when determining the transition rates.
\Cref{fig:petersen}A shows the edge state transition rates for the first eight generalized Petersen graphs $P_{n,k}$ \citep{Watkins1969_JCombTheo}, averaged within edge equivalence classes (\cref{fig:petersen}B), at a representative choice of parameter values which we fix henceforth to focus on the effects of network topology. Inspection of \cref{fig:petersen}A reveals that there are some graphs, such as $P_{6,2}$, which exhibit distinct classes obeying near-identical average rates, despite these edges' differing global symmetries. These turn out to be edges with similar sizes of cycles running through them.

When $\mu \gg \lambda$, state transitions will conserve flux throughout and so take the form of adding or subtracting a unit of flux around an entire Eulerian subgraph $\Gamma' \subset \Gamma$. The energy barrier to such a transition increases with the number of edges $m$ in $\Gamma'$. Indeed, suppose the transition consists of flipping a fraction $\rho$ of the edges in $\Gamma'$ from  $\phi = 0$ to  $\phi = \pm 1$, with the remaining edges necessarily flipping from $\phi = \pm 1$ to $\phi = 0$. The transition can then be approximated by a one-dimensional reaction coordinate $s$ running from $0$ to $1$, as follows. Suppose that only edges in $\Gamma'$ change during the transition (which is approximately true for $K_4$; SI Energy Barriers). Using the symmetry of~$V$, the energy $H(s)$ at point $s$ of the transition is given by $H(s) = H_0 + \rho m V(s) + (1-\rho)m V(1-s)$ for $H_0$ a constant dependent on the states of the edges not in~$\Gamma'$. The energy barrier is then $\Delta H = \max_s H(s) - H(0)$. $H$ is maximized precisely when $\rho V(s) + (1-\rho)V(1-s)$ is maximized, which is independent of $m$. Therefore, for fixed~$\rho$, $\Delta H$ is linear in $m$. 
This argument suggests that, since the transition rate $k \propto \exp(-\beta\Delta H)$, edges contained in small cycles should have exponentially greater transition rates than those with longer minimal cycles.
Define the \emph{$e$-girth}~$g_e$ to be the minimum length of all cycles containing edge~$e$, so that the usual graph girth is $\min_e g_e$.
Categorizing edge classes in \cref{fig:petersen} by $g_e$ confirms our hypothesis: the transition rates divide into near-distinct ranges where larger $g_e$ yields rarer transitions, and  equivalence classes with similar rates have identical $e$-girths.

\subsection*{Asymmetric networks}
Even for graphs with no symmetry, the behavior of each edge can still be predicted by a simple local heuristic. For our purposes, a graph with `no symmetry' is one possessing only the identity automorphism, in which case we say it is \emph{asymmetric}~\citep{GodsilRoyle}.
In this case, edges can have transition rates entirely distinct from one another.
\Cref{fig:asymmetric}A depicts the mean transition rates for the edges of 20 asymmetric bridgeless cubic graphs on 21 edges \Methods, exemplified in SI Movie 2.
As in \cref{fig:petersen}, categorizing edges by their $e$-girths (illustrated in \cref{fig:asymmetric}B for the starred graph; see Fig.~S4 for all 20 graphs) splits the rates into near-distinct bands, despite the total absence of symmetry. However, the bands are not perfectly distinct, and high-girth edges in particular display a range of transition rates both within and across graphs. A large portion of this variation is accounted for by considering the sizes of \emph{all} cycles containing an edge.
While the full dependence is highly complex, we can obtain a good transition rate estimate by considering just two cycles. 
Let $\ell_1 = g_e$ and $\ell_2$ be the sizes of the two smallest cycles through~$e$. (It may be that $\ell_1 = \ell_2$.) Drawing on our earlier argument for the transition rate of an $m$-cycle, suppose that flips of these two cycles occur independently with waiting times $T_i$ distributed exponentially at rates $\Lambda_i = \gamma\exp(-\alpha\ell_i)$ for constants $\alpha,\gamma$. The waiting time~$T = \min\{T_1,T_2\}$ for one of these to occur is then exponentially distributed with rate $\Lambda_1 + \Lambda_2$. Therefore, $\langle T \rangle = (\Lambda_1 + \Lambda_2)^{-1}$ and so the transition rate $k = 1/\langle T \rangle = \gamma G$, where we have defined the \emph{girth-weighted rate}
\begin{align}
\label{eq:GWM}
G = \exp(-\alpha \ell_1) + \exp(-\alpha \ell_2).
\end{align}
Fitting $k = \gamma G$ to the data in \cref{fig:asymmetric}A yields an exponent $\alpha = 1.31$.
This gives a strong match to the data (\cref{fig:asymmetric}C): the different $e$-girth categories now spread out along the fit line, showing that \cref{eq:GWM} yields an easily computed heuristic to estimate the transition rates of edges in a given graph better than $g_e$ alone.

\begin{figure}[b!]
\centering
\includegraphics[width=\columnwidth]{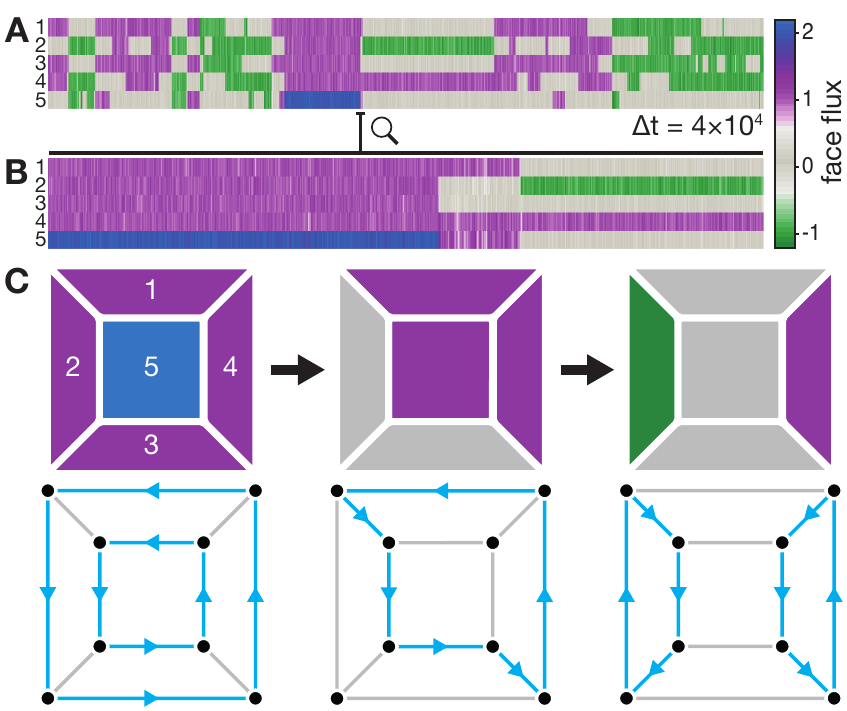}
\caption{Incompressible flow on planar graphs can be represented using a face-based cycle basis.
(A)~Flux--time traces for flow about each of the internal faces of $P_{4,1}$, as labelled in (C), from \cref{eq:langevin_F} with $\lambda = 2.5$, $\beta^{-1} = 0.05$.
(B)~Zoom of trace showing a transition between two 8-cycles, which are global minima, via a 6-cycle.
(C)~Distinct state configurations in (B) of face fluxes (upper) and corresponding edge flows (lower).
\label{fig:faceswitching}}
\end{figure}

\section*{Discussion}

\subsection*{Incompressible limit}
Thus far, we have been considering approximate incompressibility with $\mu \gg \lambda$ but finite. We now pass to the fully incompressible limit $\mu \rightarrow \infty$, which necessitates a change of flow representation.
In this limit, the dynamics of $\Phi$ are constrained to the null space $\ker \tD$, and so $\Phi$ must be decomposed using a basis of $\ker \tD$ (SI Incompressible Limit). The most physically intuitive decomposition uses a \emph{cycle basis}
comprising a non-orthogonal set of unit-flux cycles, so that each basis element corresponds to adding or removing a unit of flux around a single cycle. Finite planar graphs in particular possess a highly intuitive cycle basis. Fix a planar embedding for $\Gamma$. Let each $\{F_\alpha\}$ be the component of anti-clockwise flux around each of the $|\cE| - |\cV| + 1$ non-external (finite) faces of $\Gamma$, and define the flux about the external (infinite) face to be zero.
The flux on an edge is then simply the difference of the fluxes about its two adjacent faces.
In particular, let $\tA = (A_{\alpha e})$ be the matrix whose rows are the cycle basis vectors, so that $\phi_e = F_\alpha A_{\alpha e} $. This implies $F_\alpha = P_{\alpha e}\phi_e$ for $\tP = (\tA \tA^\transpose)^{-1} \tA$. The components $F_\alpha$ then obey
\begin{align}
\label{eq:langevin_F}
dF_\alpha =  -(\tP \tP^\transpose)_{\alpha\gamma} \frac{\partial \hat H}{\partial F_\gamma} dt + \sqrt{2\beta^{-1}} dX_{\alpha,t},
\end{align}
where $\hat H$ is the reduced energy $\hat H = \lambda \sum_{e \in \cE} V(F_\alpha A_{\alpha e})$,
and~$\mathbf{X}_t$ is a vector of correlated Brownian noise with covariance matrix $\tP \tP^\transpose= (\tA \tA^\transpose)^{-1}$.
Now, $A_{\alpha e}$ is non-zero only when edge $e$ borders face $\alpha$, and is then $+1$ or $-1$ depending on the orientation of the edge relative to the face. Therefore, $\tA$ is all but one row of the incidence matrix of the planar dual of $\Gamma$, where the missing row is that corresponding to the external face, meaning $\tilde\tL = \tA \tA^\transpose$ is the Laplacian on vertices of the dual (its \emph{Kirchhoff matrix}) with the row and column corresponding to the external face deleted. Thus the independent edge noise turns into correlated noise with covariance~$\tilde\tL^{-1}$ which is typically non-zero almost everywhere. In other words, flux conservation means that the noise on one edge is felt across the entire graph.

\subsection*{Example} 
\Cref{fig:faceswitching} shows an integration of \cref{eq:langevin_F} for an embedding of the graph $P_{4,1}$ (\cref{fig:petersen}), the cube,
whose covariance matrix $\tilde\tL^{-1}$ is non-zero everywhere (SI Incompressible Limit).
(In fact, the dual of a polyhedral graph is unique~\citep{Whitney1932_AmJMath}.)
Note that the $F_\alpha$ need not only fluctuate around states $\{-1,0,1\}$, as seen in \cref{fig:faceswitching} when a state with $F_5 = +2$ is attained. The constraint now is that the \emph{difference} $F_\alpha - F_\beta$ between adjacent faces $\alpha$ and $\beta$ must be near $\{-1,0,1\}$, as this is the flux on the shared edge. Here, the central face $F_5$ can  assume $\pm 2$ if its neighbors are all $\pm1$. In general, a face of minimum distance $d$ to the external face, which is constrained to zero flux, can be metastable at values up to $\pm d$ if all its neighbors are at $\pm(d-1)$. 
A further example on a $15 \times 15$ hexagonal lattice is given in Fig.~S5.

\subsection*{Low temperature limit and ice-type models} 
 Similar to how a lattice $\phi^4$ theory generalizes the Ising model \citep{Wioland2016_NatPhys}, on a regular lattice our model in the incompressible limit gives a lattice field theory generalization of ice-type or loop models~\citep{Baxter,Kondev1997_PRL,Batchelor1994_PRL}. Instead of there being a finite set of permitted flow configurations at each vertex, we now have a continuous space of configurations. Taking the low temperature limit $\beta\lambda \rightarrow \infty$ then recovers a discrete vertex model with $\phi_i \in \{-1,0,1\}$, where allowed configurations must be maximally flowing; thus, for example, a square lattice yields the six-vertex ice model~\citep{Baxter}.
For general $\Gamma$, the $\beta\lambda \rightarrow \infty$ limit can be understood as a form of random subgraph model \citep{Grimmett},
where the ground states are flows on maximum Eulerian subgraphs which are selected uniformly with a multiplicity of two for either orientation of every sub-cycle.
On a cubic graph, a subset of the ground states are the Hamiltonian cycles (cycles covering every vertex exactly once), if they exist, since a maximally-flowing state will have two out of every three edges at every vertex flowing. The expected number of Hamiltonian cycles on a cubic graph grows like $|\cV|^{-1/2}(4/3)^{|\cV|/2}$ as $|\cV| \rightarrow \infty$~\citep{Robinson1992_RandStructAlg}, meaning large cubic graphs  possess a huge number of ground states.

\subsection*{Complex networks}
We have focused on small regular graphs, but the dynamical principles presented here will still apply to active flow on complex networks.
The edges in a large random graph typically exhibit a wide distribution of $e$-girths, where topologically protected edges, whose $e$-girth is large enough to prevent them ever changing state within a realistic observation window, coexist with frequently switching edges of small $e$-girth.
In fact, graphs drawn from distributions modeling real-life network phenomena \citep{Watts1998_Nature,Barabasi1999_Science} seem to have far more small $e$-girth edges than their fixed degree or uniformly random counterparts (SI Complex Networks).
Furthermore, though large random graphs are almost always asymmetric~\citep{GodsilRoyle}, many real-life complex networks have very large automorphism groups \citep{MacArthur2008_DAM} meaning that, as in \cref{fig:petersen}, there will be large sets of edges in such a network with identical transition rates.
Active flow on complex networks can therefore be expected to display a rich phenomenology of local and global state transitions.

\section*{Conclusions}

Our analysis shows that the state transition statistics of actively driven quasi-incompressible flow networks can be understood by combining reaction rate theory with graph-theoretic symmetry considerations.  Furthermore, our results suggest that non-equilibrium flow networks may offer new insights into ice-type models and \textit{vice versa}. The framework developed here offers ample opportunity for future generalizations both from a biophysical and a transport optimization perspective. For example, an interesting open biological question concerns how plasmodial organisms such as \textit{Physarum}~\cite{Nakagaki2000_Nature,Tero2010_Science,Alim2013_PNAS} adapt and optimize their network structure in response to external stimuli, such as light or nutrient sources or geometric constraints~\citep{Reid2012_PNAS,Alim2013_CurrBiol}. Our investigation suggests that a combined experimental and mathematical analysis of cycle structure may help explain the decentralized computation strategies employed by these organisms.  More generally, it will be interesting to explore whether similar symmetry-based statistical approaches can guide the topological  optimization of other classes of non-equilibrium networks, including neuronal and man-made information flow networks that typically operate far from equilibrium.

\matmethods{\Cref{eq:langevin,eq:langevin_F} were integrated by the Euler--Maruyama method~\citep{Higham2001_SIAMRev} with time step $\delta t = 5\times 10^{-3}$. Mathematica (Wolfram Research, Inc.) was used to generate and analyze all graphs. For full details, see SI Numerical Methods. All data available on request.}

\showmatmethods

\acknow{We thank R.~Goldstein, I.~Pivotto and G.~Royle for discussions.
This work was supported by Trinity College, Cambridge (F.G.W.), NSF Award CBET-1510768 (A.F. and J.D.), ARC Discovery Grant DP130100106 (J.B.F.), an MIT Solomon Buchsbaum Fund Award (J.D.), and an Alfred P. Sloan Research Fellowship (J.D.).}

\showacknow


\end{document}